\documentclass[aps, 11pt, onecolumn,
  floatfix, fleqn, notitlepage,
  amsmath, amssymb, amsfonts,
  preprintnumbers, showpacs, showkeys]{revtex4-1}\usepackage[]{graphicx}\usepackage[]{color}
\makeatletter
\def\maxwidth{ %
  \ifdim\Gin@nat@width>\linewidth
    \linewidth
  \else
    \Gin@nat@width
  \fi
}
\makeatother

\definecolor{fgcolor}{rgb}{0.345, 0.345, 0.345}

\usepackage{framed}
\makeatletter
 {\par\unskip\endMakeFramed%
 \at@end@of@kframe}
\makeatother

\definecolor{shadecolor}{rgb}{.97, .97, .97}
\definecolor{messagecolor}{rgb}{0, 0, 0}
\definecolor{warningcolor}{rgb}{1, 0, 1}
\definecolor{errorcolor}{rgb}{1, 0, 0}
\newenvironment{knitrout}{}{} 

\usepackage{alltt}

\usepackage[margin=1.25in]{geometry}
\usepackage[colorlinks=true, linkcolor=blue, citecolor=blue,
            filecolor=blue, urlcolor=blue]{hyperref}
\usepackage{dcolumn, graphicx, moreverb}
\newcolumntype{.}{D{.}{.}{-1}} 
\usepackage[american]{babel}

\usepackage{mathptmx}
\usepackage[T1]{fontenc}

\newcommand{\be}{\begin{equation}}
\newcommand{\ee}{\end{equation}}
\newcommand{\bea}{\begin{eqnarray}}
\newcommand{\eea}{\end{eqnarray}}

\newcommand{\q}[2]{\ensuremath{#1\ \mathrm{#2}}} 
\newcommand{\vol}[1]{\textbf{#1}} 

\newcommand{\trev}{\ensuremath{T_{\mathrm{rev}}}}
\newcommand{\frev}{\ensuremath{f_{\mathrm{rev}}}}
\IfFileExists{upquote.sty}{\usepackage{upquote}}{}
\begin{document}

\title{Notes on the design of experiments and beam diagnostics with
  synchrotron light detected by a gated photomultiplier for the
  Fermilab superconducting electron linac and for the Integrable
  Optics Test Accelerator (IOTA)}

\author{G.~Stancari} \email[Email:]{$\langle$stancari@fnal.gov$\rangle$.}

\author{A.~Romanov}
\author{J.~Ruan}
\author{J.~Santucci}
\author{R.~Thurman-Keup}
\author{A.~Valishev}

\affiliation{Fermi National Accelerator Laboratory, PO Box 500,
  Batavia, Illinois 60510, USA\footnote[1]{This manuscript has been
    authored by Fermi Research Alliance, LLC under Contract
    No.~DE-AC02-07CH11359 with the U.S.\ Department of Energy, Office
    of Science, Office of High Energy Physics.}}

\date{\today}

\begin{abstract}
  We outline the design of beam experiments for the electron linac at
  the Fermilab Accelerator Science and Technology (FAST) facility and
  for the Integrable Optics Test Accelerator (IOTA), based on
  synchrotron light emitted by the electrons in bend dipoles, detected
  with gated microchannel-plate photomultipliers (MCP-PMTs). The
  system can be used both for beam diagnostics (e.g., beam intensity
  with full dynamic range, turn-by-turn beam vibrations, etc.) and for
  scientific experiments, such as the direct observation of the time
  structure of the radiation emitted by single electrons in a storage
  ring. The similarity between photon pulses and spectrum at the
  downstream end of the electron linac and in the IOTA ring allows one
  to test the apparatus during commissioning of the linac.
\end{abstract}

\preprint{FERMILAB-FN-1043-AD-APC}

\maketitle



\section{Introduction}

Synchrotron radiation has been widely used as non-destructive beam
diagnostic~\cite{Hofmann:2004}. We propose to use a gated
photomultiplier to collect synchrotron light generated by the beam in
bending dipoles, both at the downstream end of the electron linac of
the Fermilab Accelerator Science and Technology (FAST) facility and in
the Integrable Optics Test Accelerator~\cite{Antipov:JINST:2016}.

The goals include both advanced beam diagnostics and scientific
experiments. The main objectives are the following:
\begin{itemize}
\item Characterize synchrotron-light signal and backgrounds in these
  specific accelerator environments;
\item Develop a bunch-by-bunch (for the linac) or turn-by-turn (for
  IOTA) intensity monitor with a wide dynamic range, from nominal
  intensities down to single electrons;
\item Record turn-by-turn beam vibrations in IOTA with high
  sensitivity, for experiments in beam dynamics;
\item Directly observe the time structure of radiation emission from a
  single electron in a storage ring.
\end{itemize}

\section{Beam pulse structure}

The linac bunches have an intensity of $3\times 10^7$~particles (5~pC) to
$2\times 10^{10}$ (3~nC), a pulse width of the order of picoseconds,
and a spacing of 333~ns (111~ns may also be possible). The pulse train
can contain between 1 and 3000~bunches (1~ms maximum pulse width),
with a repetition rate of 1~Hz.

IOTA will circulate single nominal bunches of $2\times 10^9$
electrons, with a revolution period of 133~ns. The rms bunch length
will be 0.4~ns. Intensities can be lowered down to single electrons in
a controlled way by manipulating the orbit, lattice or radiofrequency
voltage.

Linac and IOTA beams will have similar pulse structures.  In both
cases, the bunch length is short compared to the time resolution of a
typical photomultiplier, so that photon signals within a bunch overlap
in the detector output.

\section{Expected signal}

\begin{table*}[b]
\begin{ruledtabular}
\caption{Chosen experimental cases and typical synchrotron-light parameters.} 
\label{tab:cases-table}
\begingroup\footnotesize
\begin{tabular}{ldddddddd}
  & \multicolumn{1}{c}{Lorentz} & \multicolumn{1}{c}{Radius} & \multicolumn{1}{c}{Critical freq.} & \multicolumn{1}{c}{Critical energy} & \multicolumn{1}{c}{Power} & \multicolumn{1}{c}{Energy loss} & \multicolumn{1}{c}{Photon flux} & \multicolumn{1}{c}{Full cone} \\ 
   & \multicolumn{1}{c}{factor $\gamma$}
                            & \multicolumn{1}{c}{$\rho$ [m]}
                            & \multicolumn{1}{c}{$\omega_c$ [$10^{15}$ rad/s]}
                            & \multicolumn{1}{c}{$E_{\gamma c}$ [eV]}
                            & \multicolumn{1}{c}{$P_s$ [nW]}
                            & \multicolumn{1}{c}{$U_s$ [eV/turn]}
                            & \multicolumn{1}{c}{$n_s$ [/turn]}
                            & \multicolumn{1}{c}{$\phi$ [mrad]} \\ \hline
D600 300 MeV &  587 & 4.89 & 18.6 & 12.2 & 0.229 &  147 & 38.9 & 3.41 \\ 
  D600 200 MeV &  391 & 4.89 & 5.51 & 3.63 & 0.0452 & 28.9 & 25.9 & 5.11 \\ 
  D600 150 MeV &  294 & 4.89 & 2.33 & 1.53 & 0.0143 & 9.16 & 19.4 & 6.81 \\ 
  D600 100 MeV &  196 & 4.89 & 0.689 & 0.454 & 0.00283 & 1.81 &   13 & 10.2 \\ 
  IOTA 150 MeV &  294 &  0.7 & 16.2 & 10.7 & 0.699 &   64 & 19.4 & 6.81 \\ 
  IOTA 100 MeV &  196 &  0.7 & 4.81 & 3.17 & 0.138 & 12.6 &   13 & 10.2 \\ 
  \end{tabular}
\endgroup
\end{ruledtabular}
\end{table*}

\begin{table*}[b]
\begin{ruledtabular}
\caption{Calculation of expected photoelectron yield for each sample case.} 
\label{tab:expected-signal}
\begingroup\footnotesize
\begin{tabular}{ldddd}
  & \multicolumn{1}{c}{Avg. Q.E.} & \multicolumn{1}{c}{Error on avg. Q.E.} & \multicolumn{1}{c}{Average number of collected} & \multicolumn{1}{c}{Integration error on $N_{pe}$} \\ 
  
  & \multicolumn{1}{c}{[$10^{-3}$]}
  & \multicolumn{1}{c}{[$10^{-5}$]}
  & \multicolumn{1}{c}{photoelectrons $N_{pe}$ [$10^{-4}$/$e^-$/pass]}
  & \multicolumn{1}{c}{[$10^{-6}$/$e^-$/pass]} \\ \hline
D600 300 MeV & 8.72 &  1.5 & 1.84 & 0.315 \\ 
  D600 200 MeV & 6.89 & 1.57 & 1.45 & 0.33 \\ 
  D600 150 MeV & 3.07 & 2.49 & 0.647 & 0.525 \\ 
  D600 100 MeV & 0.108 & 1.37 & 0.0228 & 0.289 \\ 
  IOTA 150 MeV & 8.74 & 1.53 & 1.84 & 0.321 \\ 
  IOTA 100 MeV & 6.38 & 1.53 & 1.34 & 0.323 \\ 
  \end{tabular}
\endgroup
\end{ruledtabular}
\end{table*}

Synchrotron light is emitted by electrons in bending dipoles,
collected through a window in the straight-through section of beam
pipe, transported through lenses and mirrors, and collected by a
photomultiplier positioned on top of the dipole magnet.

We calculate the expected number of photoelectrons per electron per
pass using classical synchrotron radiation formulas for long
magnets~\cite{Hofmann:2004}, the transmission of the optical components,
and the spectral response of the photomultiplier.

The layout of the accelerator is described in detail in
Ref.~\cite{Antipov:JINST:2016}. A few cases with different energies are chosen
(Table~\ref{tab:cases-table}), both for the D600 15-degree dipole at
the downstream end of the 300-MeV beam line of the electron linac and
the 30- and 60-degree dipoles in IOTA (150-MeV nominal electron
energy).

\begin{knitrout}\small
\definecolor{shadecolor}{rgb}{0.969, 0.969, 0.969}\color{fgcolor}\begin{figure}
\includegraphics[width=\textwidth]{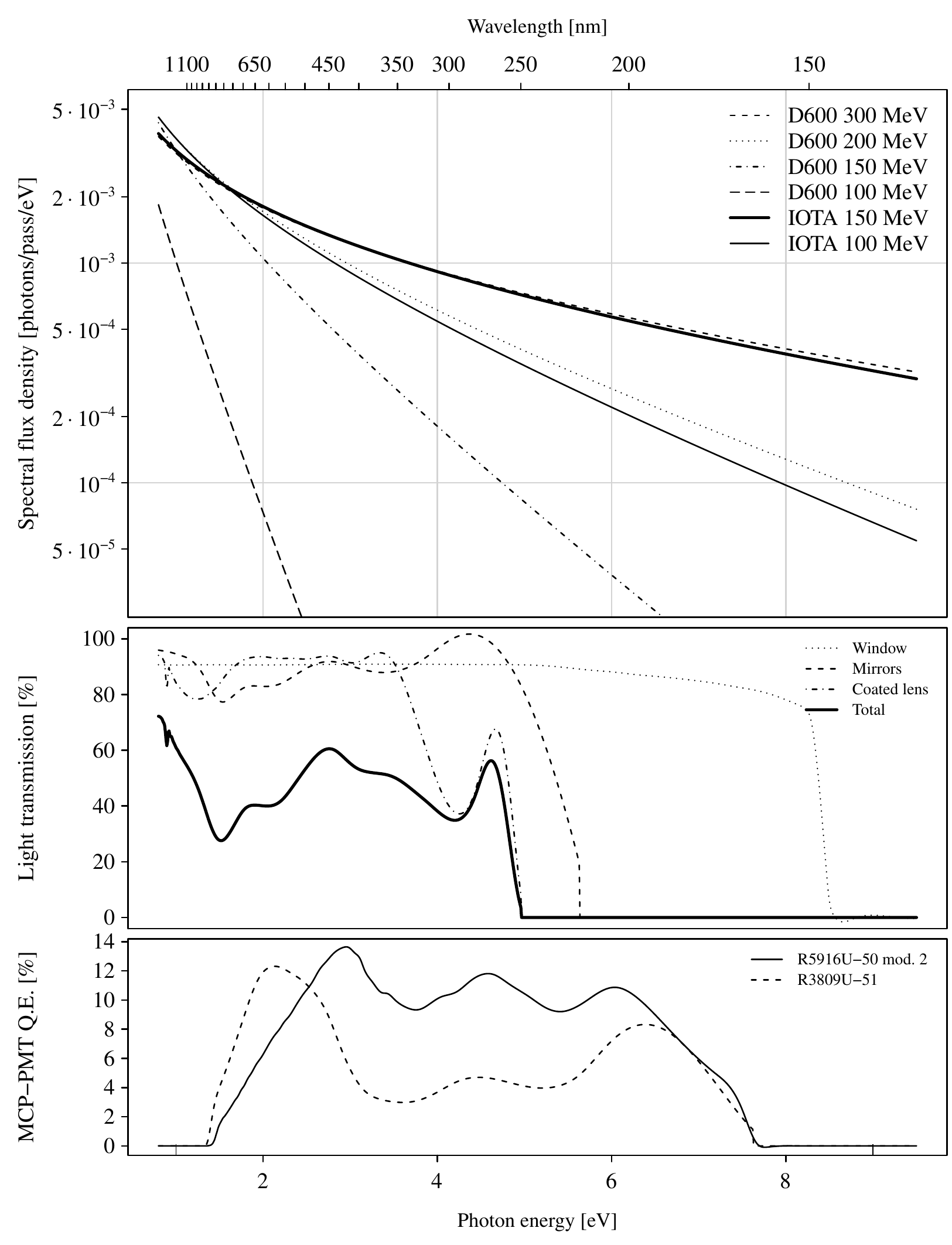} \caption[Spectral flux density of synchrotron-light photons (top)]{Spectral flux density of synchrotron-light photons (top); transmission of light-transport system (middle); quantum efficiency of photodetector (bottom).}\label{fig:spectra}
\end{figure}

\end{knitrout}

The spectral photon fluxes are shown in Figure~\ref{fig:spectra} for
different cases. With the appropriate choice of beam energy, spectra
and fluxes at the D600 dipole can be made remarkably similar to those
expected at the IOTA dipoles.

The light transport system is assumed to collect the full
synchrotron-radiation cone. It includes a 6-mm-thick crystalline quartz window,
4 UV-enhanced aluminum mirrors, and a fused silica lens with UV antiriflective coating. The
transmission of each individual element (from the manufacturer) and
the total transmission are shown in Figure~\ref{fig:spectra} (middle
plot) as a function of photon energy and wavelength.

The quantum efficiency of the photodetectors is also shown in
Figure~\ref{fig:spectra} (bottom plot). Two curves are drawn: one for
the existing MCP-PMT from the Tevatron Synclite
system~\cite{Thurman-Keup:Beams-doc-1975:2008} (Hamamatsu R5916U-50 mod.~2) and
one for a commercial infrared-enhanced device
(Hamamatsu R3809U-51). (Although its infrared response is enhanced, the
loss of quantum efficiency over the rest of the spectrum yields a
lower overall signal.)

The average number of photons per electron per pass is obtained by
multiplying the quantum efficiency, averaged over the emission and
transmission spectra, by the total number of photons per pass. The
results are shown in Table~\ref{tab:expected-signal}.

For a single electron circulating in IOTA (revolution period
$\trev = \q{ 133}{ns}$, revolution
frequency $\frev = \q{7.52}{MHz}$),
we expect an average counting rate of
\q{1.4}{kHz},
with a negligible probability of 2~photons being emitted in the same
dipole. A small number of circulating electrons, and discrete steps in
counting rates due to losses of electrons, should be clearly
detectable with counting times of the order of 1~s. For nominal IOTA
intensities, the analog output of the photomultiplier can be
integrated to provide a signal proportional to the beam
current. Reading out the photomultiplier signal in both counting and
current mode allows one to cover the full dynamic range, from
1~electron to nominal bunches.

For preliminary experiments at the D600 dipole, one can test the
current readout mode with nominal IOTA bunches (300~pC) and the
counting mode with low-intensity bunches (e.g., 3~pC) and a neutral
density filter to suppress multiple photons from the same bunch.

A continuously variable neutral-density filter with high gradient can
be used to translate beam displacements into intensity modulation of
synchrotron light. Similar experiments were done in the past using a
hard-edge screen~\cite{Nesterenko:HEACC:2001}, but beam alignment was
time-consuming. The variable neutral density filter should be less
sensitive to beam alignment, at the cost of a poorer spatial
sensitivity.

The ability to store and monitor single electrons in IOTA opens up
several research areas~\cite{Riehle:NIMA:1988, Pinayev:NIMA:1994,
  Aleshaev:NIMA:1995, Pinayev:NIMA:1996, Klein:Metrologia:2010}. One
of them is the study of the physics of synchrotron radiation and its
effect on the dynamics of single particles. In particular, we are
interested in the direct observation of the time series of photon
emissions of a single particle in a storage ring: is it random,
chaotic, or regular? This study can be started with the initial IOTA
configuration, and enhanced at a later stage by the insertion of an
undulator (a possible synergy with the optical stochastic cooling
experiments).

\section{Backgrounds and systematic effects}

One of the main purposes of preliminary studies at the D600 dipole
location is to assess background levels and their
fluctuations. Typical background and noise sources include thermionic emission,
radiation (natural or accelerator-induced), afterpulsing, and light
leaks.

A gated photodetector provides strong background reduction. Another
technique we will investigate is a lengthening of the
synchrotron-radiation path in order to delay the signal pulse with
respect to the arrival of the beam.

The effect of the magnetic field on the photomultiplier response needs
to be investigated as well.

\section{Apparatus}
\label{sec:apparatus}

\subsection{Light collection and transport}

The light transport system is
designed to collect the full synchrotron-radiation cone emitted in the
body of the dipole magnet (D600 in the linac, 30-deg or 60-deg in
IOTA).

Radiation emitted by the electron beam in the dipole leaves the vacuum
chamber through a 6-mm-thick crystalline quartz window. Transport and focusing is
provided by 4 UV-enhanced aluminum mirrors and a
fused silica lens with UV antiriflective coating. A variable neutral-density filter controls the
intensity of the transmitted light. The microchannel-plate photomultiplier
(Hamamatsu R5916U-50 mod.~2) is placed in a dark box on top of the dipole
magnet, together with other diagnostics, such as the
synchrotron-radiation cameras for beam position and size measurement.

\subsection{Data acquisition}

\begin{figure}[b]
  \centering
  \includegraphics[width=\textwidth]{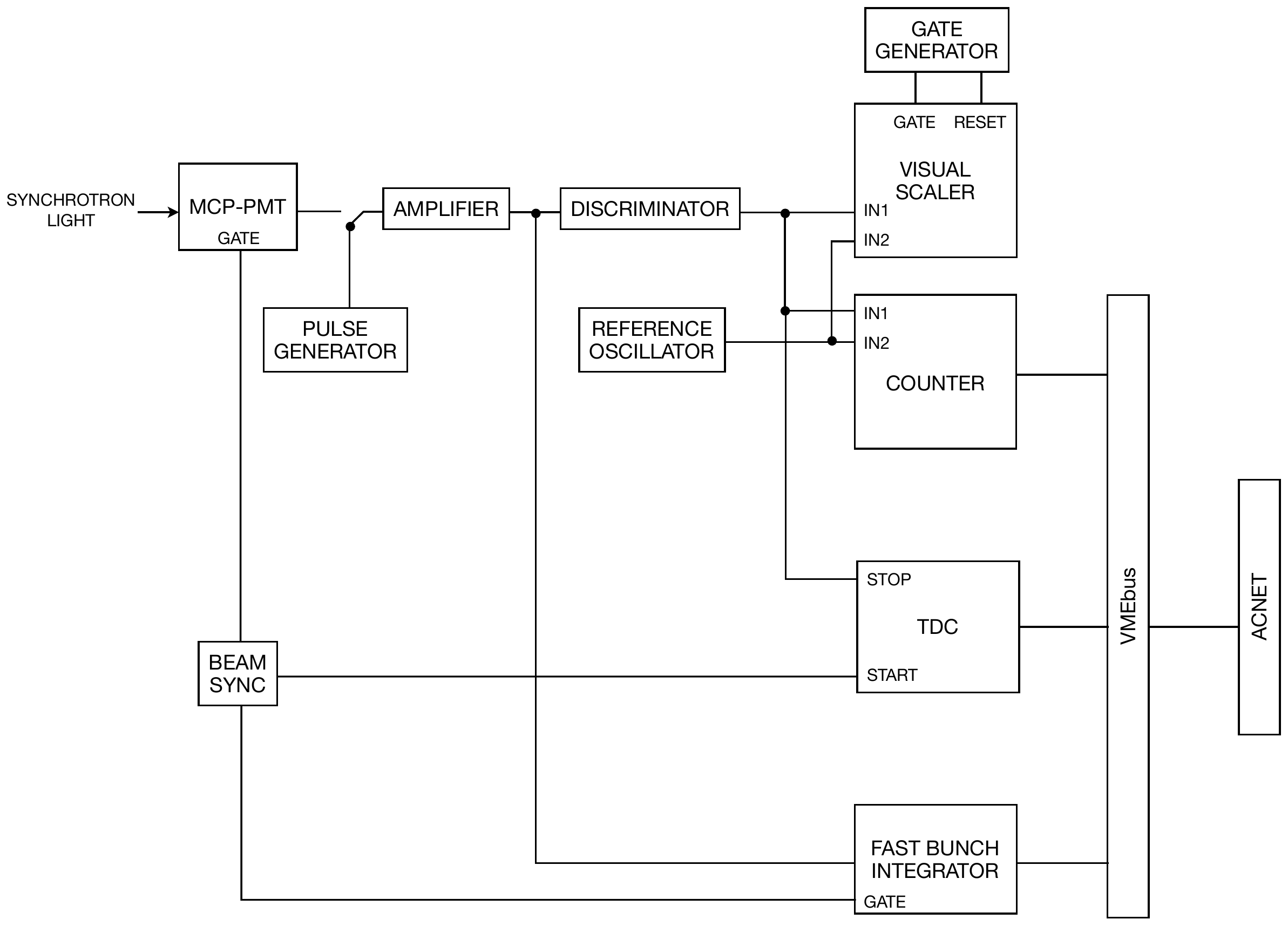}
  \caption{Schematic diagram of the data-acquisition system.}
  \label{fig:daq}
\end{figure}

The data-acquisition system (Figure~\ref{fig:daq}) is located in the
Electrical Service Building (ESB) above the IOTA enclosure.

The signal from the photomultiplier is split into~3 readout chains:
1~analog for pulse-height analysis and 2~digital for counting and timing.

The analog chain consists of a gated fast bunch integrator (similar to
the ones used in the Tevatron and Main Injector). Individual pulses
are integrated over a time window and read out via VMEbus into ACNET
devices.

The digital chains start with a constant-fraction discriminator. The
signal is then sent to counters (a visual scaler for local readout and
a Struck SIS-3805 VME scaler) and to a time-to-digital converter
(TDC). The counting rates are available via ACNET devices.

The time-to-digital conversion is under development. It is designed to
store the time difference between the beam synchronization signal
(laser pulse in the linac, radiofrequency reference in IOTA) and the
synchrotron radiation signal for a given number of turns (for IOTA) or
microbunches (for the linac).

\section{Conclusions}

We present the design of a detection system for synchrotron
radiation with gated photomultipliers in the Fermilab Integrable
Optics Test Accelerator and electron linac for a set of beam
energies. The system can be used for both diagnostics and physics experiments.

The time structure of the beam and the spectral photon fluxes at the
downstream end of the linac and in the IOTA bending dipoles are
similar. This allows one to set up and study the system during linac
commissioning as IOTA is being constructed. Taking into account beam
energy, dipole bend radius, transmission of the optical system, and
quantum efficiency of the photodetector, the average photoelectron
yield for a typical case (IOTA 150 MeV) is
$1.8\times 10^{-4}$ for
each electron in the beam per pass in a dipole magnet
(Table~\ref{tab:expected-signal}). As a consequence, if background
levels are low (or if their fluctuations are small) compared to the
signal, intensities and losses of individual electrons should be
detectable in IOTA with integration times of the order of a second.

The addition of precise signal timing with state-of-the-art accuracy
($< \q{10}{ps}$)~\cite{Christiansen:picoTDC} would be highly valuable
to enable direct observation of the time structure of synchrotron
radiation emission from single particles in a storage ring.

\begin{acknowledgments}
  Several people are contributing to this project. In particular, we
  would like to thank M.~Andorf (NIU), K.~Carlson, B.~Fellenz,
  W.~Johnson (Fermilab), and the whole IOTA/FAST Department within the
  Accelerator Physics Center of the Fermilab Accelerator Division.
\end{acknowledgments}






\end{document}